\documentclass{iaus}
\usepackage{graphicx}
\usepackage{natbib}

\title[Physics/chemistry of HMCs] 
{Physics and chemistry of hot molecular cores}

\author[H. Beuther]   
{H. Beuther}
\affiliation{Max-Planck-Institute for Astronomy, K\"onigstuhl 17, 69117 
Heidelberg, Germany \break email: beuther@mpia.de}

\pubyear{2006}
\volume{237}  
\pagerange{1--16}
\date{?? and in revised form ??}
\jname{Triggered Star Formation in a Turbulent ISM}
\editors{B. G. Elmegreen \& J. Palous, eds.}
\begin{document}
\def\aj{AJ}%
\def\araa{ARA\&A}%
\def\apj{ApJ}%
\def\apjl{ApJ}%
\def\apjs{ApJS}%
\def\ao{Appl.~Opt.}%
\def\apss{Ap\&SS}%
\def\aap{A\&A}%
\def\aapr{A\&A~Rev.}%
\def\aaps{A\&AS}%
\def\azh{AZh}%
\def\baas{BAAS}%
\def\jrasc{JRASC}%
\def\memras{MmRAS}%
\def\mnras{MNRAS}%
\def\pra{Phys.~Rev.~A}%
\def\prb{Phys.~Rev.~B}%
\def\prc{Phys.~Rev.~C}%
\def\prd{Phys.~Rev.~D}%
\def\pre{Phys.~Rev.~E}%
\def\prl{Phys.~Rev.~Lett.}%
\def\pasp{PASP}%
\def\pasj{PASJ}%
\def\qjras{QJRAS}%
\def\skytel{S\&T}%
\def\solphys{Sol.~Phys.}%
\def\sovast{Soviet~Ast.}%
\def\ssr{Space~Sci.~Rev.}%
\def\zap{ZAp}%
\def\nat{Nature}%
\def\iaucirc{IAU~Circ.}%
\def\aplett{Astrophys.~Lett.}%
\def\apspr{Astrophys.~Space~Phys.~Res.}%
\def\bain{Bull.~Astron.~Inst.~Netherlands}%
\def\fcp{Fund.~Cosmic~Phys.}%
\def\gca{Geochim.~Cosmochim.~Acta}%
\def\grl{Geophys.~Res.~Lett.}%
\def\jcp{J.~Chem.~Phys.}%
\def\jgr{J.~Geophys.~Res.}%
\def\jqsrt{J.~Quant.~Spec.~Radiat.~Transf.}%
\def\memsai{Mem.~Soc.~Astron.~Italiana}%
\def\nphysa{Nucl.~Phys.~A}%
\def\physrep{Phys.~Rep.}%
\def\physscr{Phys.~Scr}%
\def\planss{Planet.~Space~Sci.}%
\def\procspie{Proc.~SPIE}%
\let\astap=\aap
\let\apjlett=\apjl
\let\apjsupp=\apjs
\let\applopt=\ao

\maketitle

\begin{abstract}
  Young massive star-forming regions are known to produce hot
  molecular gas cores (HMCs) with a rich chemistry. While this
  chemistry is interesting in itself, it also allows to investigate
  important physical parameters. I will present recent results
  obtained with high-angular-resolution interferometers disentangling
  the small-scale structure and complexity of various molecular gas
  components. Early attempts to develop a chemical evolutionary
  sequence are discussed. Furthermore, I will outline the difficulty
  to isolate the right molecular lines capable to unambiguously trace
  potential massive accretion disks.  
\end{abstract}

\firstsection 
\section{Introduction}

Hot molecular cores (HMCs) are characterized by gas temperatures
exceeding 100\,K and a rich chemistry observable in molecular line
emission at (sub)mm wavelength. These HMCs are considered to represent an
early evolutionary stage in high-mass star formation where the
protostars are still actively accreting and ultracompact H{\sc ii}
regions have not yet formed (e.g., \citealt{kurtz2000,beuther2006b}).
Single-dish observations toward HMCs revealed stunning molecular line
forests, but they were not capable to spatially resolve the various
molecular components (e.g.,
\citealt{blake1987,schilke1997b,hatchell1998b}). Only interferometric
high-spatial-resolution observations resolve the spatial complexity in
more detail (e.g., \citealt{wright1996,blake1996,wyrowski1999}). For
the closest and best known HMC Orion-KL, recent observations with the
Submillimeter Array (SMA) dissected its molecular components showing
significant spatial differences between, e.g, SiO, oxygen-bearing
species like CH$_3$OH, nitrogen-bearing species like CH$_3$CN or
sulphur-bearing species like SO$_2$ \citep{beuther2005a}. In the
following, I will present recent SMA results toward the HMC in G29.96
as well as a molecular comparison of high-spatial-resolution
observation of various massive star-forming regions. Finally, the
difficulties of identifying molecular line tracers for massive disks
will be discussed.

\section{The hot molecular core G29.96}

We used the SMA to observe the well-known HMC G29.96 in a broad range
of spectral line and continuum emission around 862\,$\mu$m (Beuther et
al.~in prep.). The achieved angular resolution is exceptional of the
order $0.3''$ for the continuum and $0.5''$ for the line emission. The
submm continuum data resolved the previously identified HMC (e.g.,
\citealt{cesaroni1994}) into four sub-sources within a projected area
of $\sim$6900(AU)$^2$. These four source comprise a proto-Trapezium
system, and assuming spherical symmetry one can estimate an
approximate protostellar density of $2\times 10^5$ protostars/pc$^3$.

\begin{figure}[htb]
\includegraphics[width=9cm,angle=-90]{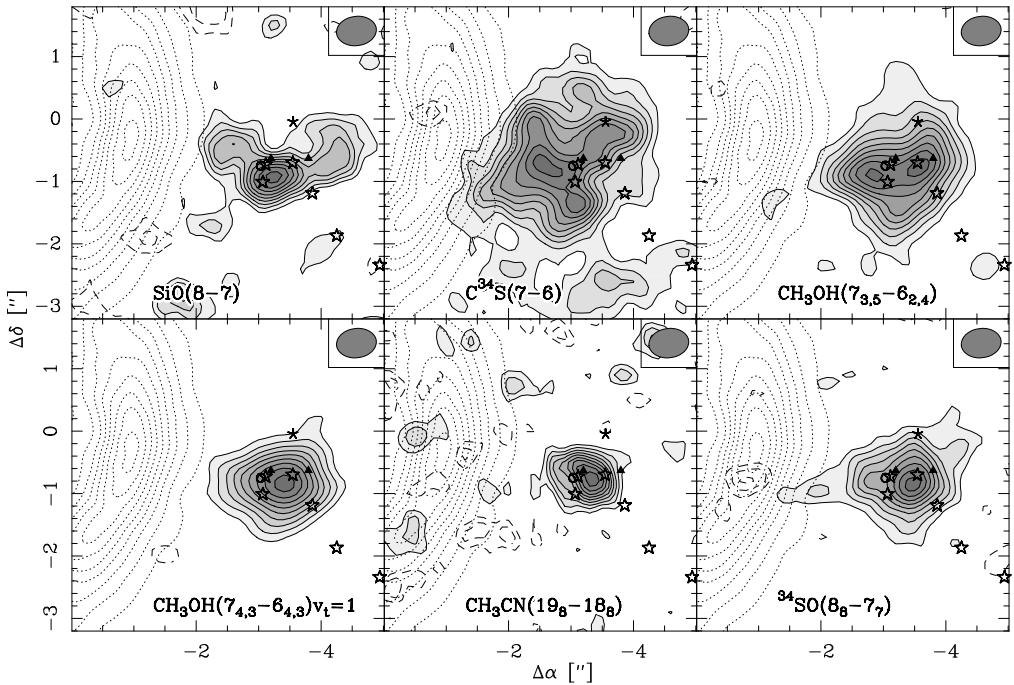}
\caption{Spectral line maps toward the HMC G29.96 observed with the
  SMA (Beuther et al.~ in prep.). The dotted contour always show the
  UCH{\sc ii} region as observed at cm wavelengths
  \citep{cesaroni1994}. The grey-scale with contours shows the
  integrated line emission from the species labeled in each panel. The
  stars mark submm continuum peaks identified during the same
  observations. The triangles and circles show H$_2$O and H$_2$CO
  maser emission, respectively \citep{hofner1996,hoffman2003}, the
  asterisks mark the mid-infrared peak from \citet{debuizer2002}, and
  the synthesized beam is shown at the top-right of each panel.}
  \label{g29_images}
\end{figure}

Within the given bandpass of 4\,GHz we detected $\sim$80 spectral
lines from 18 molecular species, isotopologues or vibrational excited
species, with a minor fraction of $\sim$5\% of unidentified lines. The
range of excitation temperatures for the set of lines varies between
40 and 750\,K, hence we are capable to study the cold and the warm gas
at the same time. Figure \ref{g29_images} presents a compilation of
integrated emission maps from a few representative species, and one
can already discern from these maps the spatial complexity of the
region. A detailed analysis of the whole dataset will be published
shortly (Beuther et al.~in prep.), here I only want to highlight a few
characteristics:\\
$\bullet$ SiO shows some extended emission. Analyzing the spectral
data-cube, we identify at least one, potentially two molecular
outflows.\\
$\bullet$ C$^{34}$S is weak toward the HMC center and the four submm
peaks, but it shows strong emission at the edge of the HMC and its
interface to the UCH{\sc ii} region. This may be interpreted as
chemical evolution: Early-on, at temperatures of the order 30\,K CS
gets released from the dust grain and the C$^{34}$S map should have
appeared centrally peaked at that time. However, when the HMC heats up
to $\geq$100\,K, H$_2$O is released from the grains, this dissociates
to OH, and the OH reacts with the S to form SO, which is centrally
peaked then (Fig.~\ref{g29_images}). This leaves significant
C$^{34}$S only at the edges of the HMC.\\
$\bullet$ No spectral line shows the same four-peaked morphology as
the submm continuum emission. Hence none traces unambiguously the
protostellar condensations. In addition to outflow contributions and
chemistry effects two other processes are considered to be important
for that. On the one hand, many spectral lines are optically thick and
therefore only trace the outer envelope of the region without
penetrating toward the central protostellar cores. On the other hand,
we are likely suffering from confusion because the molecules are not
exclusively found in the central protostellar cores but also in the
surrounding envelope. Disentangling these components is a
difficult task.\\
$\bullet$ Of the many molecular lines, only a single one exhibits a
coherent velocity structure with a velocity gradient perpendicular to
the main outflow. Since this structure comprises three of the submm
peaks it likely is a larger-scale rotating toroid which may (or may
not) harbor accretion disks closer to the protostellar condensations.
For the difficulties of massive disk studies see \S \ref{disks}.

\section{Toward a chemical evolutionary sequence}

\begin{figure}[htb]
\includegraphics[angle=-90,width=6.7cm]{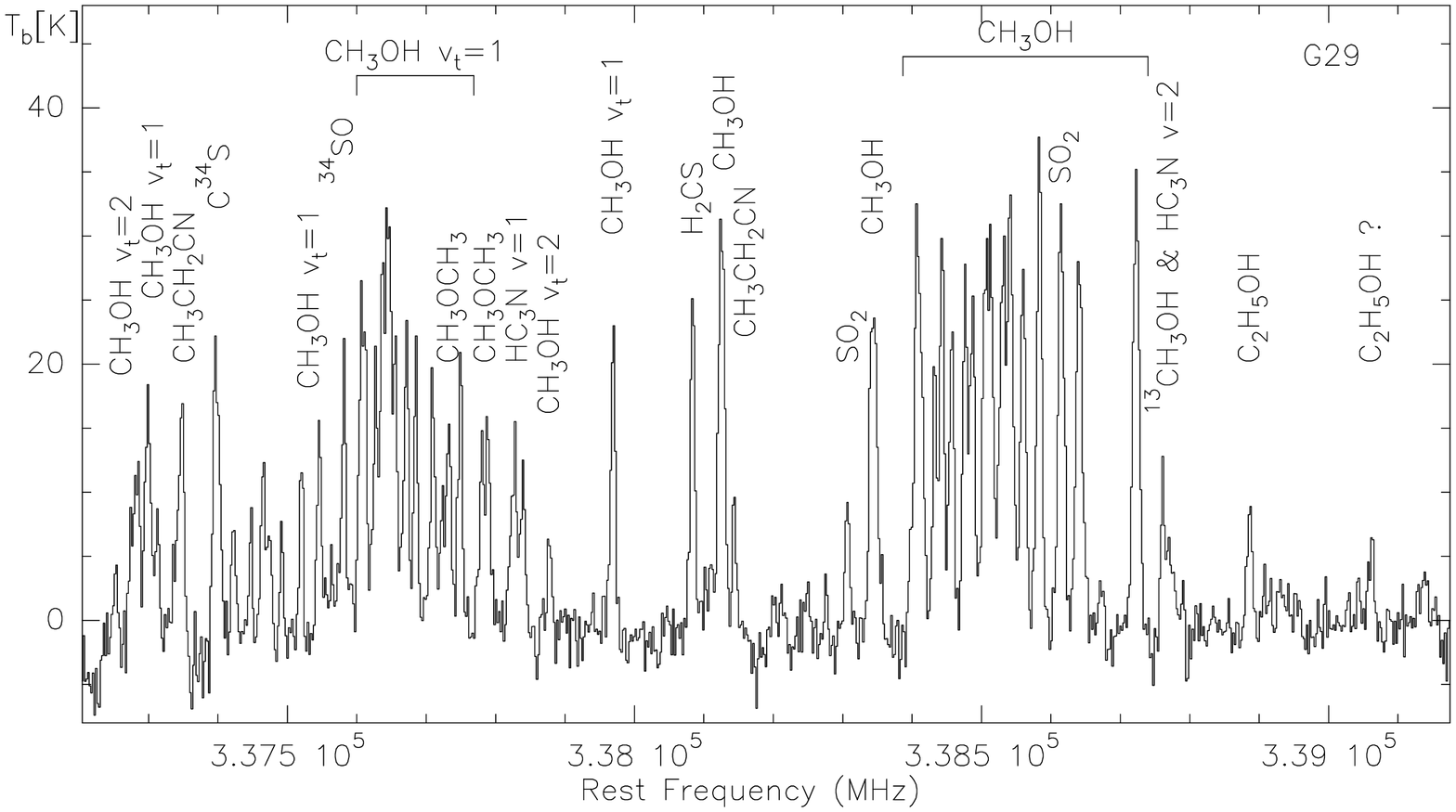}
\includegraphics[angle=-90,width=6.7cm]{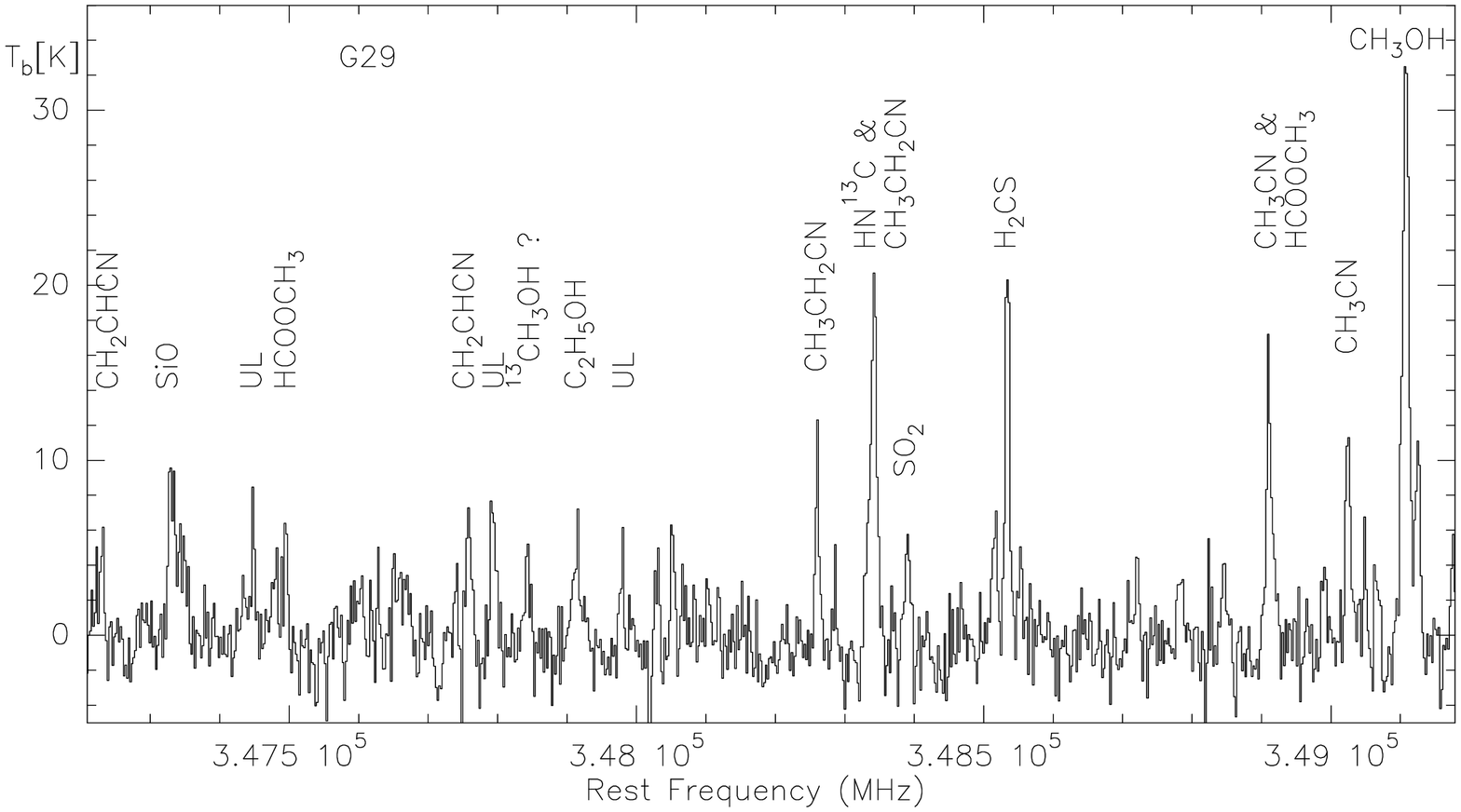}\\
\includegraphics[angle=-90,width=6.7cm]{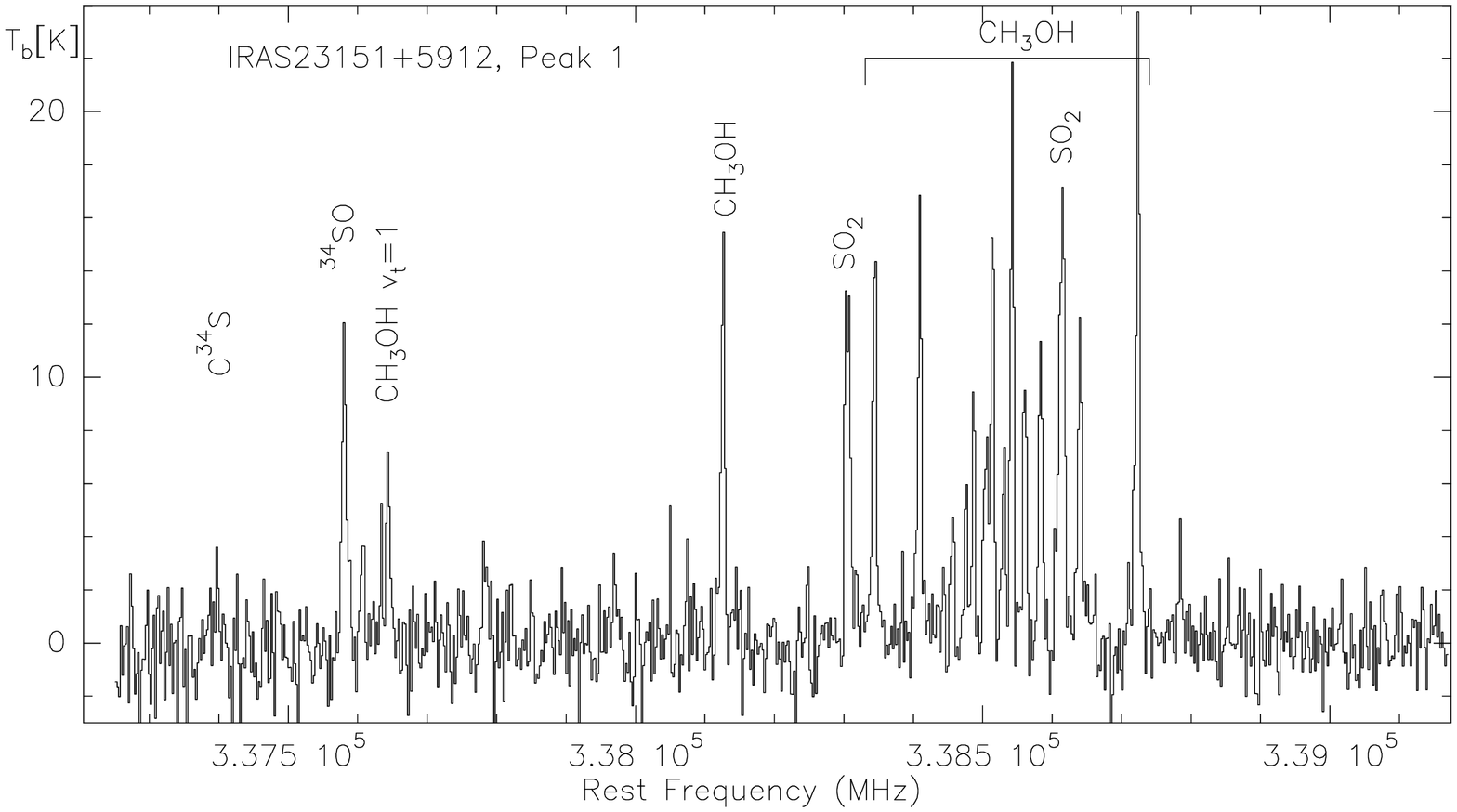}
\includegraphics[angle=-90,width=6.7cm]{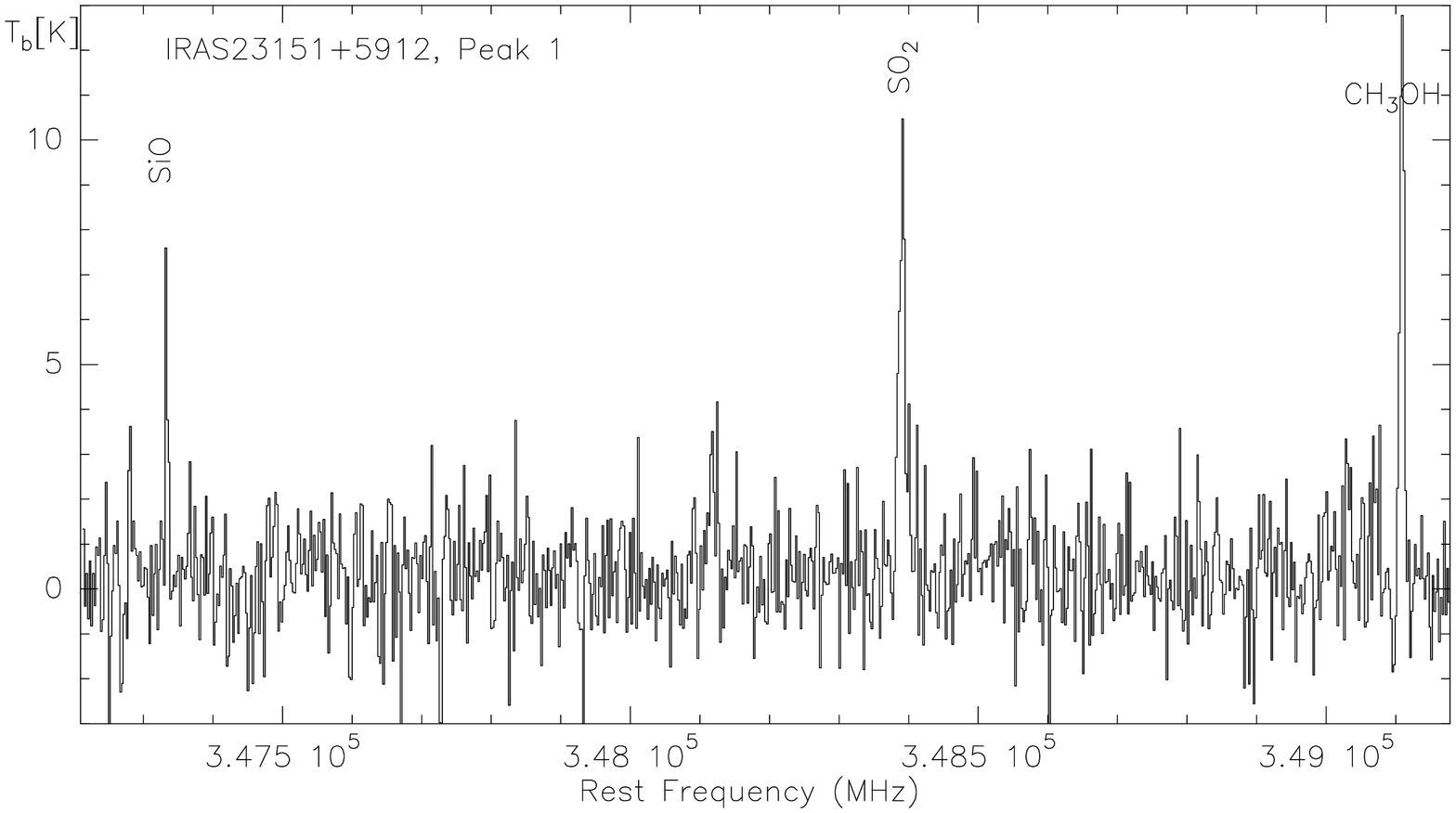}\\
\caption{SMA spectra extracted from the final data-cubes toward two
  massive star-forming regions (G29.96 top row \& IRAS\,23151+5912
  bottom row). The spectral resolution in all spectra is 2\,km/s. The
  left and right column show the lower and upper sideband data,
  respectively.}
\label{sample_spectra}
\end{figure}

With the long-term goal in mind to establish chemical sequences -- in
an evolutionary sense as well as with varying luminosity -- over the
last few years we observed four massive star-forming regions with the
SMA in exactly the same spectral setup around 862\,$\mu$m as used
originally for the Orion-KL observations \citep{beuther2005a}. These
four regions comprise a range of luminosities between
$10^{3.8}$\,L$_{\odot}$ and $10^5$\,L$_{\odot}$, and they cover
different evolutionary stages from young pre-HMCs to typical HMCs
(Orion-KL: HMC, $L\sim 10^5$\,L$_{\odot}$, $D\sim 0.45$\,kpc; G29.96:
HMC, $L\sim 9\times 10^4$\,L$_{\odot}$, $D\sim 6$\,kpc; IRAS\,23151,
pre-HMC, $L\sim 10^5$\,L$_{\odot}$, $D\sim 5.7$\,kpc; IRAS\,05358:
pre-HMC, $L\sim 10^{3.8}$\,L$_{\odot}$, $D\sim 1.8$\,kpc). Smoothing
all datasets to the same linear spatial resolution, we are now capable
to start comparing these different regions. Figure
\ref{sample_spectra} presents typical spectra extracted toward the HMC
G29.96 and the pre-HMC IRAS\,23151 (Beuther et al.~in prep.).

A detailed comparison between the four sources will be given in a
forthcoming paper (Beuther et al.~in prep.), here I just outline a
few differences in a qualitative manner.\\
$\bullet$ The HMCs show far more molecular lines than the pre-HMCs.
Orion-KL and G29.96 appear similar indicating that the nature
of the two sources may be similar as well.  Regarding the two
pre-HMCs, the higher luminosity one (IRAS\,23151) shows still more
lines than the lower-luminosity source (IRAS\,05358). Since
IRAS\,05358 is approximately three times closer to us than
IRAS\,23151, this is not a sensitivity issue but it is likely due to
the different luminosity objects forming at the core centers.\\
$\bullet$ The ground-state CH$_3$OH lines are detected toward all four
sources.  However, the vibrational-torsional excited CH$_3$OH are only
strongly detected toward the HMCs Orion-KL and G29.96. Independent of
the luminosity, the pre-HMCs exhibit only one CH$_3$OH $v_t=1$ line,
which can easily be explained by the lower average temperatures of the
pre-HMCs.\\
$\bullet$ A more subtle difference can be discerned by comparing the
SO$_2$ and the HN$^{13}$C lines near 348.35\,GHz (in the upper
sideband). While the SO$_2$ line is found toward all four sources, the
HN$^{13}$C is strongly detected toward the HMCs, but it is not found
toward the pre-HMCs. In the framework of warming up HMCs, this
indicates that nitrogen-bearing molecules are either released from the
grains only at higher temperatures, or they are daughter molecules
which need some time during the warm-up phase to be produced in
gas-phase chemistry networks. In both cases, such molecules are
expected to be found not much prior to the formation of a detectable
HMC.

\section{Identifying molecules for massive disk studies}
\label{disks}

\begin{figure}[htb]
\includegraphics[angle=-90,width=4.5cm]{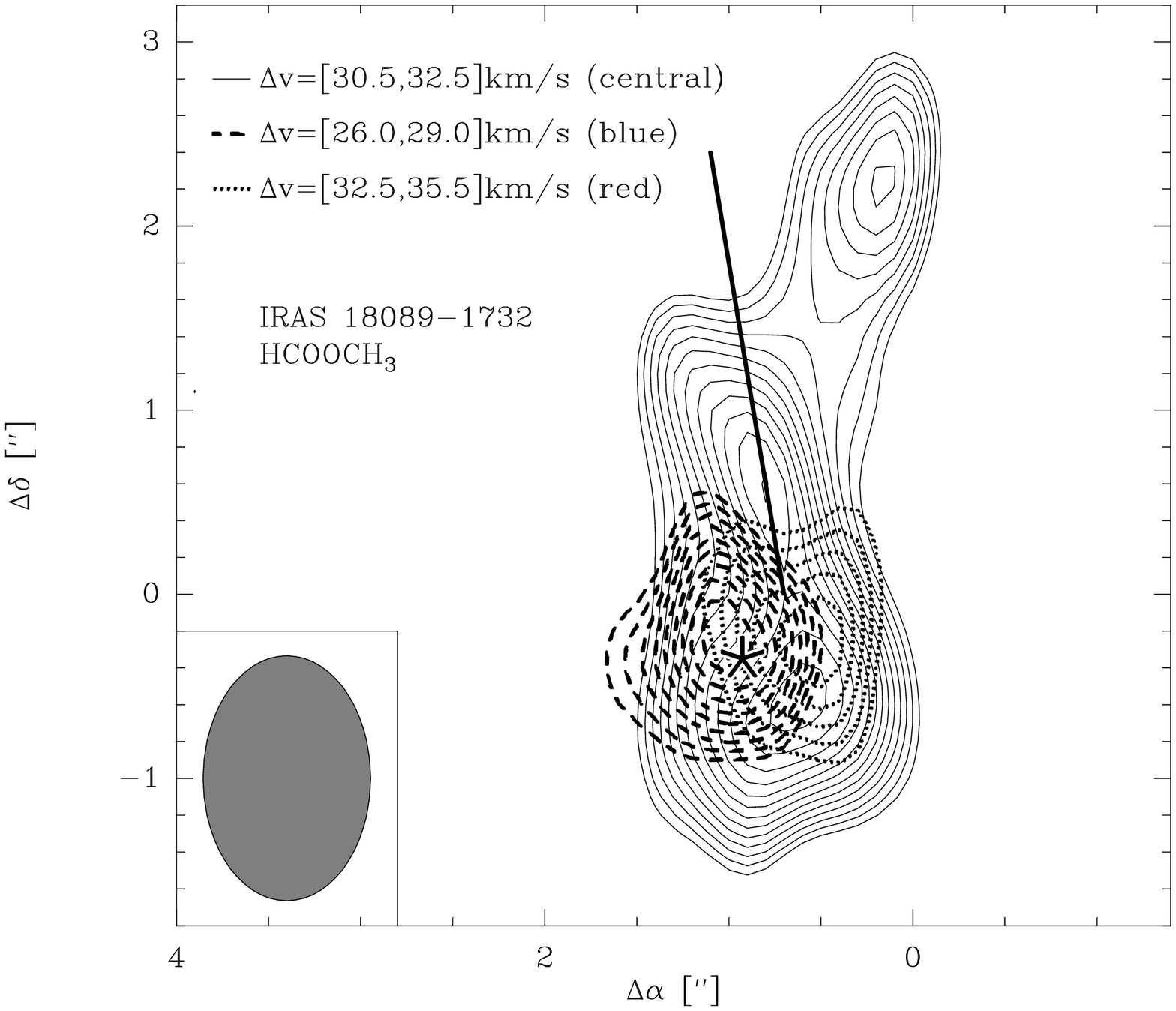}
\includegraphics[angle=-90,width=4.0cm]{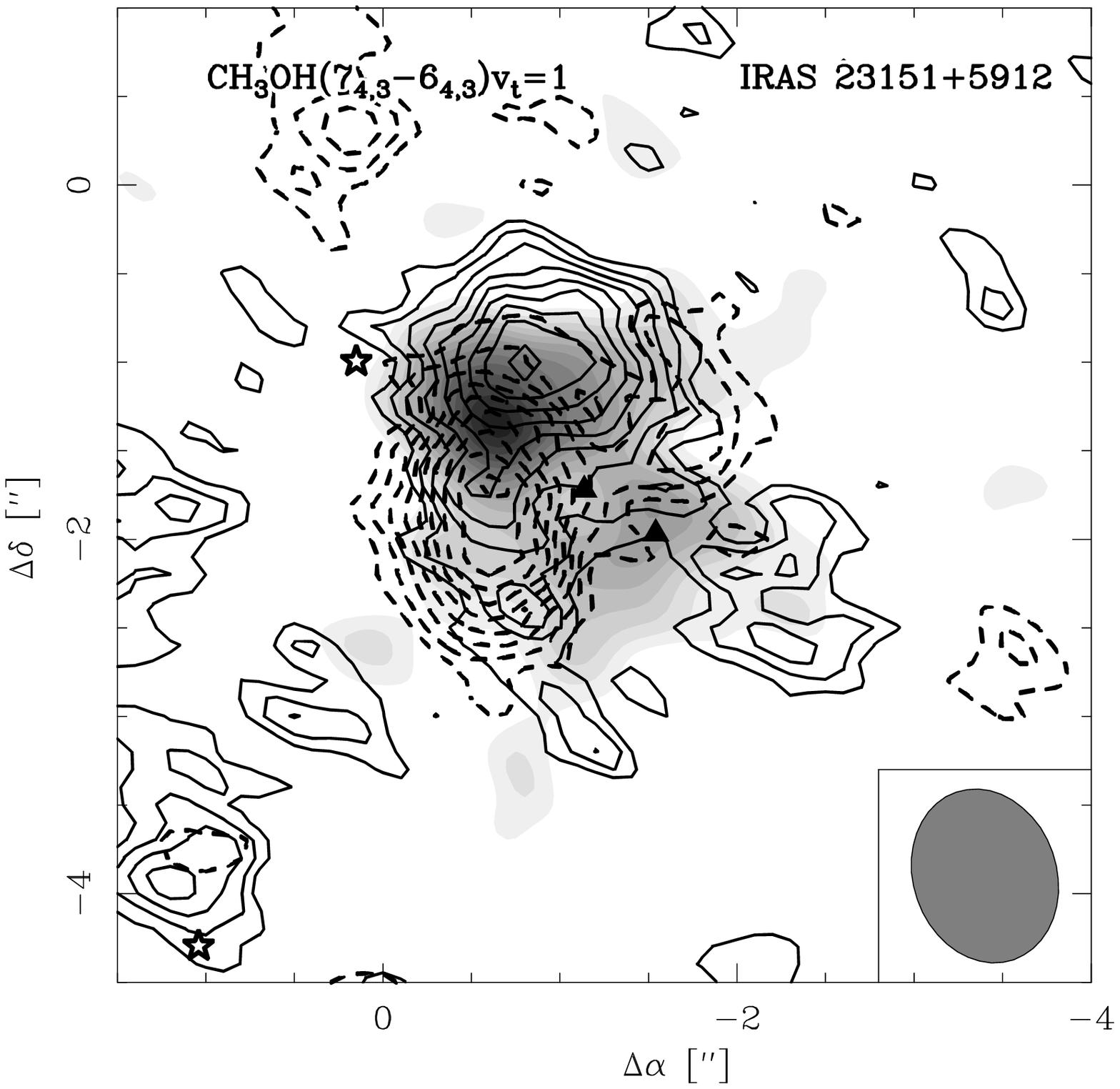}
\includegraphics[angle=-90,width=4.7cm]{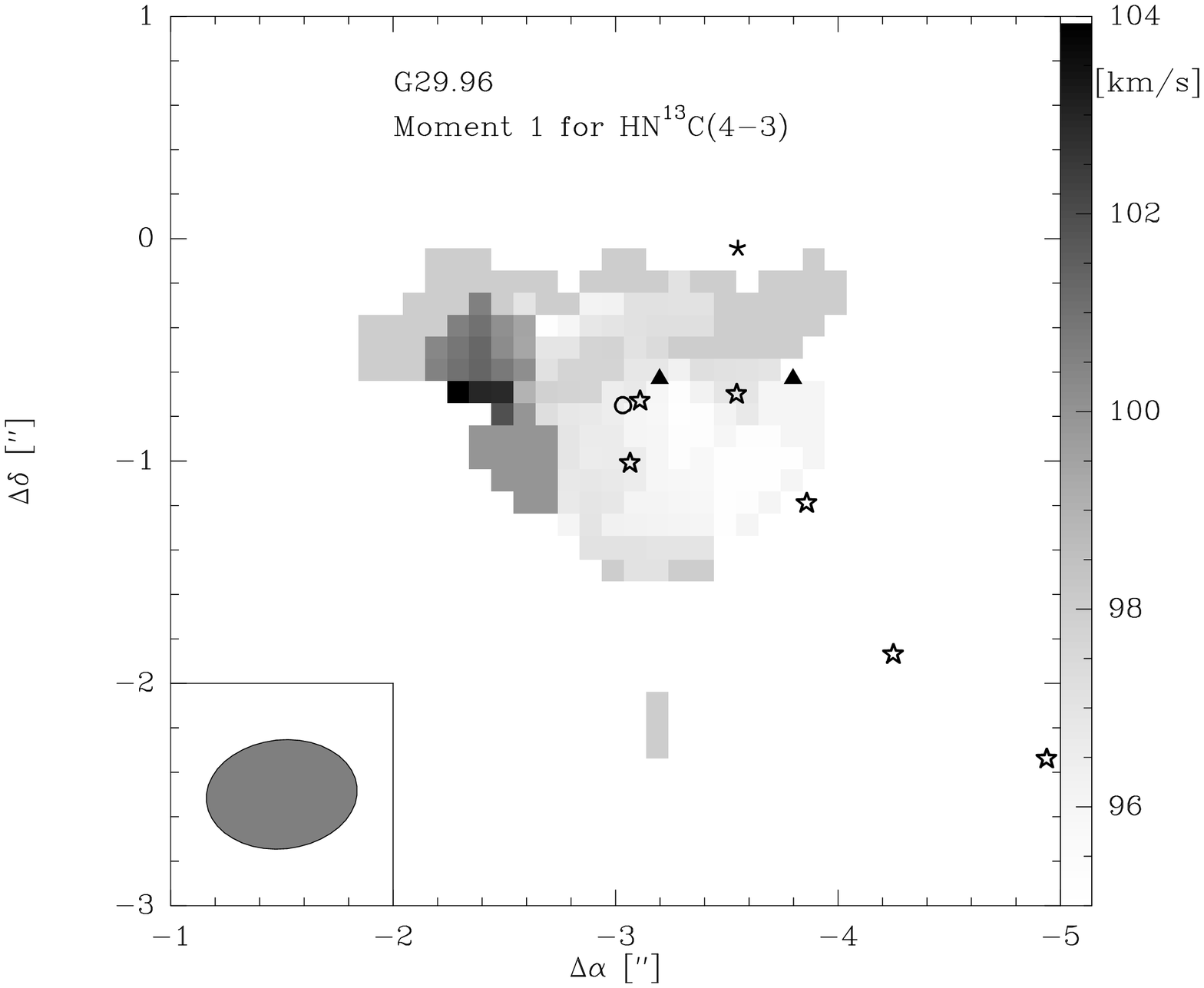}\\
\includegraphics[angle=-90,width=6.1cm]{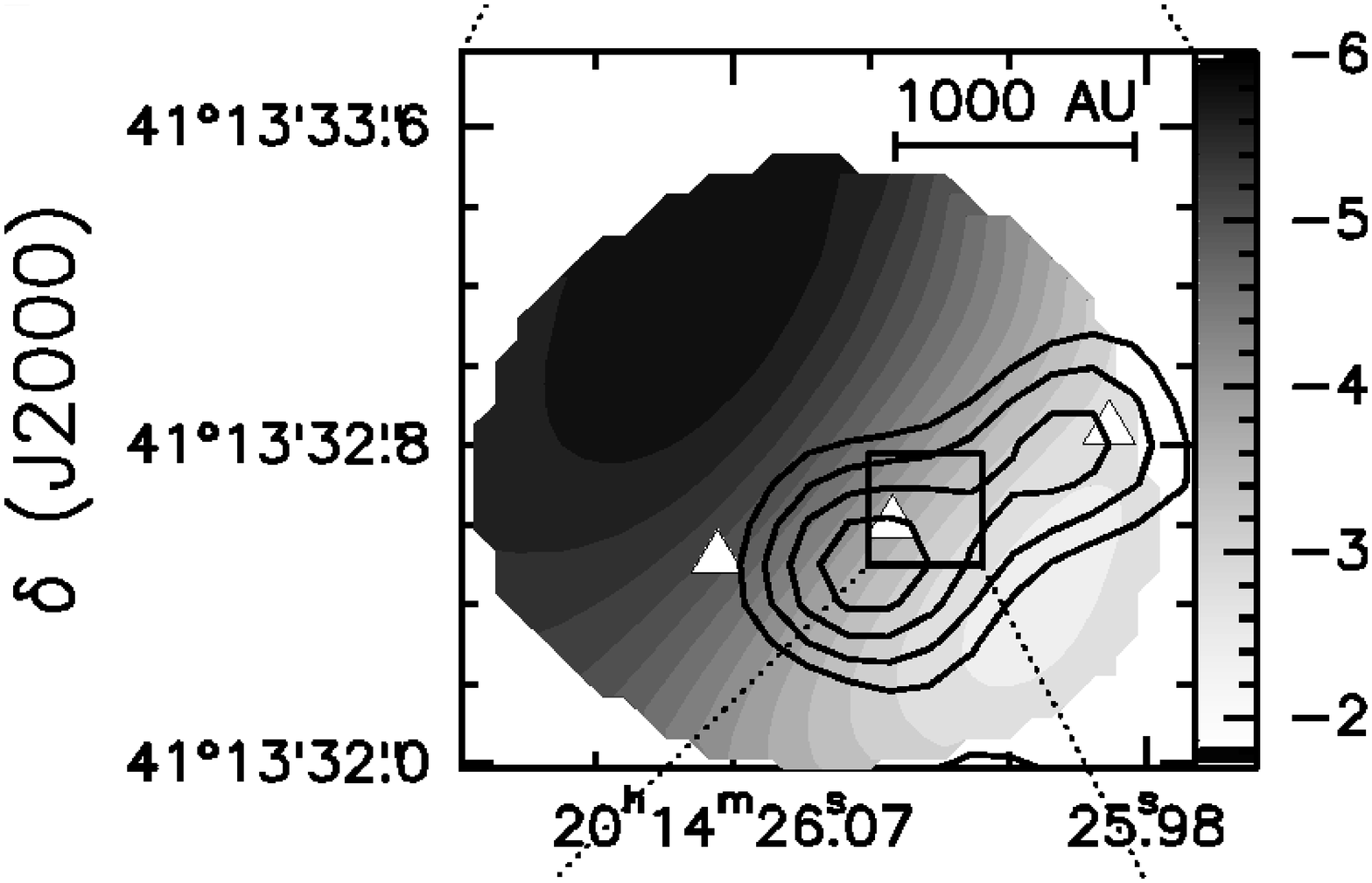}
\includegraphics[angle=-90,width=5.0cm]{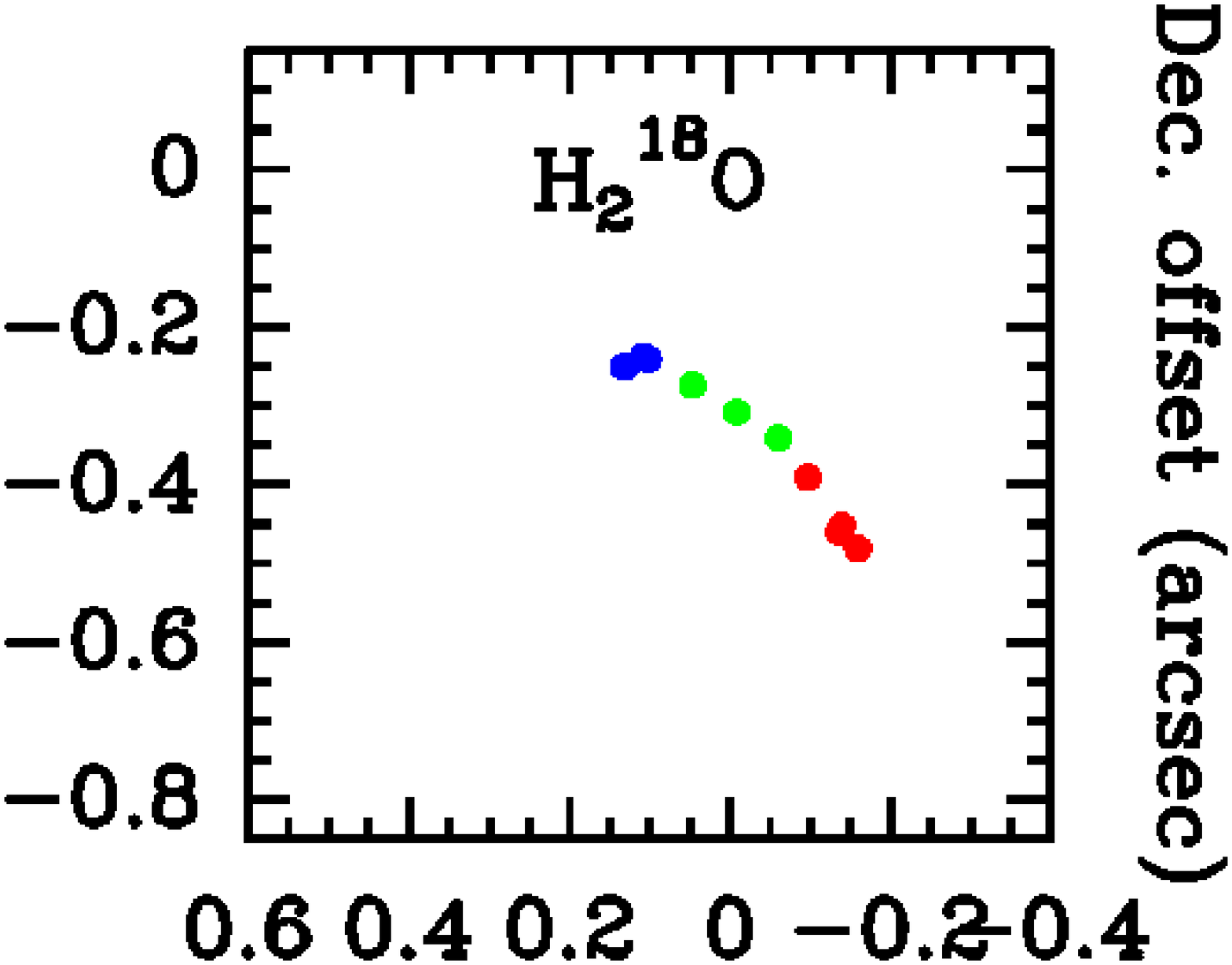}
\caption{Examples of rotation-tracing molecules: top-left: HCOOCH$_3$
  in IRAS\,18089-1732 \citep{beuther2005c}, top-middle: CH$_3$OH
  $v_t=1$ in IRAS\,23151+5912 (Beuther et al.~in prep.), top-right:
  HN$^{13}$C in G29.96 (Beuther et al.~in prep.), bottom-left:
  C$^{34}$S (and also CH$_3$CN) in IRAS\,20126+4104
  \citep{cesaroni1999,cesaroni2005}, bottom-right: H$_2^{18}$O in
  AFGL2591 \citep{vandertak2006}.}
\label{disk_examples}
\end{figure}

There exists ample, however indirect evidence for the existence of
massive disks in high-mass star formation (e.g.,
\citealt{cesaroni2006,beuther2006b}). Theorists predict that massive
accretion disks have to exist (e.g.,
\citealt{jijina1996,yorke2002,krumholz2006a}), and the observations of
collimated jet-like molecular outflows from at least B0 stars indicate
the presence of underlying accretion disks as well (e.g.,
\citealt{beuther2005b,arce2006}). However, we have not found much
observational evidence for massive accretion disks, not to speak that
we have not characterized them properly yet \citep{cesaroni2006}.  The
best known example is the disk in IRAS\,20126+4104 which even shows a
Keplerian velocity profile, but the mass of the central object is only
$\sim$7\,M$_{\odot}$, and it is probably still in its accretion phase
\citep{cesaroni2005}. There are more sources observed where we find
rotational signatures in the central cores perpendicular to the
molecular outflows (see Figure \ref{disk_examples} for a small
compilation), however, the velocity structure is not Keplerian or they
are that large (potentially comprising several sub-sources) that they
rather resemble larger-scale rotating toroids than typical accretion
disks (e.g., \citealt{cesaroni2006,keto2006}). Maybe these sources
harbor genuine accretion disks at their very centers.

A major observational problem arises because it is difficult to
disentangle the spectral line distributions from the various gas
components (mainly core-disk, envelope and outflow). Some previously
believed good disk-tracers have been shown to be strongly influenced
by the molecular outflows (e.g., CN \citealt{beuther2004e}, HCN Zhang
priv.\,comm.). Probably even more difficult, the expected central disks
and the close-by surrounding envelopes are both warming up quickly due
to the heating of the central accreting source. Therefore, the
chemical properties -- and hence the molecular emission arising from
both components -- are similar. This way, one may have to model
always both components together. Furthermore, as we have seen in the
previous sections, the chemistry varies with evolution. For example,
while C$^{34}$S may be a potentially good disk tracer in young
pre-HMCs like IRAS\,20126+4104 (Fig.~\ref{disk_examples}), it
obviously does not work in more evolved HMCs like G29.96
(Fig.~\ref{g29_images}). In contrast, nitrogen-bearing molecules like
HN$^{13}$C appear to be a good tracer of rotation in the HMC G29.96,
but it remains undetectable in younger sources like IRAS\,23151+5912
(Fig.~\ref{sample_spectra}).  An additional complication arises from
varying optical depths: while the 1\,mm lines of CH$_3$CN are good
rotation tracers in some sources \citep{cesaroni1999,beltran2004}, the
more excited lines in the submm bands do not always show these
signatures. \citet{beuther2005c} interpreted this difference due to
increased optical depth at the given high temperatures in the submm
bands.

The advent of broad spectral bandpasses in new or upgraded
interferometers now allows to observe many spectral lines
simultaneously (Fig.~\ref{sample_spectra}). Thus, we can identify the
best rotation-tracing molecules for individual sources after the
observations without the need of strong molecular pre-selection
effects. While one would like to observe a large sample of massive
disk candidates in the same spectral line to study the kinematic
properties as consistently as possible, this may be impossible due to
the above discussed physical and chemical difficulties. However, if we
are able to observe many sources systematically in a spectral setup
covering the most important molecules, we can select the adequate line
for each source and still investigate a larger sample in a
statistically consistent manner.

\section{Conclusions and outlook}

Interferometry at (sub)mm wavelengths is the tool of choice if one
wants to disentangle the chemical and physical complexity in massive
star-forming regions. Only since a couple of years we are capable to
spatially map a broad range of molecular lines in selected regions at
high-spatial resolution. 

One of the first results of these studies is that the spatial
diversity of the molecules is extremely complex. For a proper
understanding of the given data, it is necessary to enhance the models
including physical properties like shocks, outflows, rotation and
heating as well as chemical networks containing gas-phase and
grain-surface reactions. Furthermore, the physical and chemical models
have to be treated with state-of-the art radiative transfer codes
to finally produce synthetic images which can be compared with the
observations (e.g., \citealt{pavlyuchenkov2006}).  On the observational
side, we need to observe larger source-samples consisting of different
evolutionary stages as well as different luminosities. Observations
and modeling have advanced very much over the last decade, but often
the various groups did not interact enough. For a better understanding
of the chemical complexity of massive star-forming regions only a
concerted effort from theory, modeling and observations is likely to
result in significant progress.

Massive accretion disks are considered the holy grail in high-mass
star formation research. The availability of broad spectral bandpasses
now allows to observe larger source samples in a less pre-selective
way since it is likely that one of the observed spectral lines will
trace the central rotating structure and hence allow a kinematic
analysis. Utilizing the currently available (sub)mm interferometers
(mainly PdBI, SMA and CARMA) as well as ALMA in the coming decade, we
are expecting to reach a much better understanding of massive
accretion disks and thus high-mass star formation in general.

\begin{acknowledgments}
  Thanks a lot to Hendrik Linz for comments on an early draft of
  this paper.  H.B.~acknowledges financial support by the
  Emmy-Noether-Program of the Deutsche Forschungsgemeinschaft (DFG,
  grant BE2578).
\end{acknowledgments}





\end{document}